\providecommand{\nothesis}[1]{#1}
\providecommand{\thesis}[1]{}

\nothesis{
\newcommand{\fullver}[1]{#1}

\providecommand{\fullver}[1]{}
\fullver{\newcommand{\shortver}[1]{}}
\providecommand{\shortver}[1]{#1}

\documentclass[oribibl]{llncs}

\fullver{\usepackage[margin=1in]{geometry}}
\usepackage{amssymb,amsmath}
\usepackage{times}
\usepackage{dsfont}
\usepackage{mlapa}
\newcommand{\hide}[1]{}

\renewcommand{\note}[1]{}
\renewcommand{\note}[1]{~\\\frame{\begin{minipage}[c]{\textwidth}\vspace{2pt}\center{#1}\vspace{2pt}\end{minipage}}\vspace{3pt}\\}

\newcommand{\One}{\mathbf{1}}
\newcommand{\Real}{\mathds{R}}
\newcommand{\Rational}{\mathds{Q}}

\title{A General Framework for Computing Optimal Correlated Equilibria in Compact Games\shortver{\\ (Extended Abstract)}}
\author{Albert Xin Jiang \and Kevin Leyton-Brown}
\institute{Department of Computer Science,\\ University of British Columbia, Vancouver, Canada\\
\email{[jiang,kevinlb]@cs.ubc.ca}}
\begin{document}
\maketitle
\begin{abstract}
We  analyze the problem of computing a correlated equilibrium that optimizes some objective (e.g., social welfare).
\emcite{PR08JACM} gave a sufficient condition for the tractability of this problem; however, this condition only applies to a subset of existing representations.
We propose a different algorithmic approach for the optimal CE problem that applies to \emph{all} compact representations, and give
a sufficient condition that generalizes that of \emcite{PR08JACM}.
In particular, we  reduce the optimal CE problem to the \emph{deviation-adjusted social welfare problem}, a combinatorial optimization problem closely related to the optimal social welfare problem.
This framework allows us to identify new classes of games for which the optimal CE problem is tractable; we show that graphical polymatrix games on tree graphs are one example.
We also study the problem of computing the optimal \emph{coarse correlated equilibrium}, a solution concept closely related to CE.
Using a similar approach we derive a sufficient condition for this problem, and use it to prove that the problem is tractable for
singleton congestion games.
\end{abstract}
}
\thesis{
\chapter{A General Framework for Computing Optimal Correlated Equilibria in Compact Games}\label{ch:optCE}
}
\section{Introduction}
\thesis{In this chapter we\footnote{This chapter is based on joint work with Kevin Leyton-Brown. A slightly shorter version was submitted for publication in August 2011 and is currently under review.} 
continue to focus on correlated equilibrium (CE).
We have seen from the previous chapter and its related literature \cite{PR08JACM,JiangLB10exact} that finding a sample CE is tractable, even for compactly represented games. However, since in general there can be
an infinite number of
CE even in a generic game, finding an arbitrary one is of limited value.
}
\nothesis{
A fundamental class of computational problems in game theory is the computation of \emph{solution concepts} of finite games.
Much recent effort in the literature has concerned the problem of computing a sample Nash equilibrium \cite{ChenDeng06,Daskalakis06,Daskalakis05,GoldbergPapa06}.
First proposed by Aumann \yrcite{aumann1974subjectivity,aumann1987correlated}, correlated equilibrium (CE) is another important solution concept.
Whereas in a mixed strategy Nash equilibrium players randomize independently, in a correlated equilibrium the players can coordinate their behavior based on signals from an intermediary.

Correlated equilibria of a game can be formulated as probability distributions over pure strategy profiles satisfying certain linear constraints. The resulting linear feasibility program has size polynomial in the size of the normal form representation of the game.
However, the size of the normal form representation grows exponentially in the number
    of players. This is problematic when games involve large numbers of players.
    Fortunately, most large games of practical interest have highly-structured payoff functions, and  thus it is possible to represent them compactly. 
A line of research thus exists to look for \emph{compact game representations} that are able to succinctly describe structured games, including work on
graphical games \cite{graphical} and action-graph games \cite{ActionGraph,AGG-full}.
But now the size of the linear feasibility program for CE can be exponential in the size of compact representation; furthermore a CE can require exponential space to specify.

The problem of computing a sample CE
was recently shown to be in polynomial time for most existing compact representations \cite{PR08JACM,JiangLB10exact}. However, since in general there can be
an infinite number of
CE in a game, finding an arbitrary one is of limited value.
}
Instead, here we focus on the problem of computing a correlated equilibrium that optimizes some objective. In particular we consider \fullver{two kinds of objectives:
(1) A linear function}\shortver{optimizing linear functions} of players' expected utilities. For example, computing the best (or worst) social welfare corresponds to  maximizing (or minimizing) the sum of players' utilities, respectively.
\fullver{(2) Max-min welfare: maximizing the utility of the worst-off player. (More generally, maximizing the minimum of a set of linear functions of players' expected utilities.)
}
We are also interested in computing optimal coarse correlated equilibrium (CCE) \cite{Hannan1957}.
\nothesis{It is known}\thesis{Recall from \autoref{sec:lit_CE}} that the empirical distribution of any no-external-regret learning dynamic converges to the set of CCE, while the empirical distribution of no-internal-regret learning dynamics converges to the set of CE\nothesis{ (see e.g. \cite{AGTBook})}. Thus,
optimal CE / CCE provide useful bounds on the social welfare of the empirical distributions of these dynamics.
\fullver{Optimal CE / CCE can also be used as bounds on optimal NE since CE and CCE are both relaxations of NE. Hence they are also useful for computing (bounds on) the price of anarchy and price of stability of a game.
}
\thesis{The problems of computing optimal CE / CCE can be formulated as linear programs with sizes polynomial in the size of normal form. However, as with the rest of the thesis,  we are interested in the case when the input is a compactly-represented game.}

We are particularly interested in the relationship between the optimal CE / CCE problems and the problem of computing the optimal social welfare outcome (i.e.
strategy profile) of the game, which is exactly the optimal social welfare CE problem without the incentive constraints.
This is an instance of a line of questions that has received much interest from the algorithmic game theory community: ``How does adding incentive constraints to an optimization problem affect its complexity?'' This question  in the mechanism design setting is perhaps one of the central questions of algorithmic mechanism design \cite{NisanRonen01amd}.
Of course, a more constrained problem can in general be computationally easier than the relaxed version of the problem.
Nevertheless, results from complexity of Nash equilibria and algorithmic mechanism design suggest that adding \emph{incentive constraints} to a problem is unlikely to decrease its computational difficulty. That is, when the optimal social welfare problem is hard, we tend also to expect that the optimal CE problem will be hard as well.
On the other hand, we are interested in the other direction: when it is the case for a class of games that the optimal social welfare problem can be efficiently computed,
can the same structure be exploited to efficiently compute the optimal CE?

\nothesis{
The seminal work on the computation of optimal CE  is \cite{PR08JACM}. This paper considered the optimal linear objective CE problem and proved that the problem is NP-hard for many representations\shortver{ including graphical games,
polymatrix games, and congestion games.}\fullver{, while tractable for a couple of representations.}}\thesis{As mentioned in \autoref{sec:lit_CE},  \emcite{PR08JACM}
 considered the optimal linear objective CE problem and proved that the problem is NP-hard for many representations, while tractable for a couple of representations. We now take a more in-depth look at this paper.} 
\fullver{In particular, the  representations shown to be NP-hard include graphical games,
polymatrix games, and congestion games.
These hardness results, although nontrivial, are not surprising:
the optimal social welfare problem is already NP-hard for these representations.
}
On the tractability side, \emcite{PR08JACM} focused on so-called ``reduced form'' representations, meaning representations for which there exist player-specific partitions of the strategy profile space into payoff-equivalent outcomes. They showed that if a particular \emph{separation problem} is polynomial-time solvable, the optimal CE problem is polynomial-time solvable as well.
Finally, they  showed that this separation problem is polynomial-time solvable for bounded-treewidth graphical games, symmetric games and anonymous games.

Perhaps most surprising and interesting is the \emph{form} of Papadimitriou and Roughgarden's 
sufficient condition for tractability:
their separation problem for an instance of a reduced-form-based representation is essentially equivalent to solving the optimal social welfare problem for an instance of that representation with the same reduced form but possibly different payoffs.
In other words, if we have a polynomial-time algorithm for the optimal social welfare problem for a reduced-form-based representation,
we can turn that into a polynomial-time algorithm for the optimal social welfare CE problem.
However, \aunpcite{PR08JACM}'s sufficient condition for tractability only applies to reduced-form-based representations.
 Their definition of reduced forms is unable to handle representations that exploit linearity of utility, and in which the structure of player $p$'s utility function may depend on the action she chose.
As a result, many representations do not fall
into this characterization, such as polymatrix games, congestion games, and action-graph games.
Although the optimal CE problems for these representations are NP-hard in general, we are interested in identifying tractable subclasses
of games, and a sufficient condition that applies to all representations would be helpful.

In this \nothesis{article}\thesis{chapter}, we propose a different algorithmic approach for the optimal CE problem that applies to \emph{all} compact representations.
By applying the ellipsoid method to the dual of the LP for optimal CE, we show that the polynomial-time solvability of what we call the \emph{deviation-adjusted social welfare problem} is a sufficient condition for the tractability of the optimal CE problem.
We also give a sufficient condition for tractability of the optimal CCE problem: the polynomial-time solvability of the \emph{coarse deviation-adjusted social welfare problem}\fullver{, which we show reduces to the deviation-adjusted social welfare problem}.
\thesis{Our algorithms are instances of the black-box approach, with the required subroutines being the  computations of the deviation-adjusted social welfare problem and the coarse deviation-adjusted social welfare problem, respectively.}
We show that for reduced-form-based representations, the deviation-adjusted social welfare problem can be reduced to the separation problem of \emcite{PR08JACM}.
Thus
the class of reduced forms for which our problem is polynomial-time solvable contains the class for which the separation problem is polynomial-time solvable.
%
More generally, we show that if a representation can be characterized by ``linear reduced forms'', i.e. player-specific linear functions over partitions, then for that representation, the deviation-adjusted social welfare problem can be reduced to the optimal social welfare problem.
As an example, we show that for graphical polymatrix games on trees, optimal CE can be computed in polynomial time. Such games are not captured by the reduced-form framework.\footnote{In a recent paper \emcite{KXL11approxCE} has independently proposed an algorithm for optimal CE in graphical polymatrix games on trees. They used a different approach that is specific to graphical games and graphical polymatrix games, and it is not obvious whether their approach can be extended to other classes of games.}
\fullver{The key feature of these representations upon which our argument relies is that the partitions for player $p$ (which characterize the structure of the utility function for $p$) do not depend on the action chosen by $p$.}


On the other hand,  representations like action-graph games and congestion games have \emph{action-specific} structure,
and as a result the deviation-adjusted social welfare problems and coarse deviation-adjusted social welfare problems on these representations are structured  differently from the corresponding optimal social welfare problems.
Nevertheless, we are able to show a polynomial-time algorithm for the optimal CCE problem on
\emph{singleton congestion games} \cite{ieong2005fac}, a subclass of congestion games.
We use a symmetrization argument to reduce the optimal CCE problem to the coarse deviation-adjusted social welfare problem with player-symmetric deviations,
which can be solved using  a dynamic-programming algorithm.
This is an example where the optimal CCE problem is tractable while the complexity of the optimal CE problem is not yet known.

\section{Problem Formulation}
\nothesis{
Consider a simultaneous-move game $G=(\mathcal{N}, \{S_p\}_{p\in \mathcal{N}}, \{u^p\}_{p\in \mathcal{N}})$, where $\mathcal{N}=\{1,\ldots,n\}$ is the set of players.
Denote a player $p$, and player $p$'s set of pure strategies (i.e., actions) $S_p$. Let $m = \max_p|S_p|$.
Denote a pure strategy profile $s=(s_1,\ldots,s_n)\in S$,  
 with $s_p$ being player $p$'s pure strategy. Denote by $S_{-p}$ the set of partial pure strategy profiles of the players other than $p$.
Let $u^p$ be the vector of player $p$'s utilities for each pure profile, denoting player $p$'s utility under pure strategy profile $s$ as $u^p_s$.
}\thesis{We follow the notation of \autoref{ch:CE}. Furthermore, let $\mathcal{N}=\{1,\ldots,n\}$ be the set of players.}
Let $w$ be the vector of social welfare for each pure profile, that is $w=\sum_{p\in \mathcal{N}} u^p$,
with $w_s$ denoting the social welfare for pure profile $s$.

Throughout the \nothesis{paper}\thesis{chapter} we assume that the game is given in a representation with \nothesis{\emph{polynomial type} \cite{Papadimitriou,PR08JACM},
i.e., that the number of players and the number of actions for each player are bounded by polynomials of the size of the representation.}\thesis{polynomial type. Unlike in \autoref{ch:CE}, here we do not assume the existence of a polynomial-time algorithm for expected utility.}

\subsection{Correlated Equilibrium}
\nothesis{
A \emph{correlated distribution} is a probability distribution over pure strategy profiles, represented by a vector $x\in \mathds{R}^M$, where $M=\prod_p |S_p|$. Then $x_s$ is the probability of pure strategy profile $s$ under the distribution $x$.

\begin{definition}
A correlated distribution $x$ is a \emph{correlated equilibrium} (CE) if it satisfies the following \emph{incentive constraints}: for each player $p$ and each pair of her actions $i,j\in S_p$,
\fullver{\begin{equation}\label{eq:incentives}}\shortver{we have $}
\sum_{s_{-p}\in S_{-p}} [u^p_{is_{-p}}- u^p_{js_{-p}}] x_{is_{-p}} \ge 0\fullver{,\end{equation}}\shortver{$,}
where the subscript ``$is_{-p}$'' (respectively ``$js_{-p}$'') denotes the pure strategy profile in which player $p$ plays $i$ (respectively $j$) and the other players play according to the partial profile $s_{-p}\in S_{-p}$.
\end{definition}
\nothesis{Intuitively, when
       a trusted intermediary draws a strategy profile $s$ from this distribution,  privately announcing to each player $p$ her own component $s_p$,
       $p$ will have no incentive to choose another strategy, assuming others follow the suggestions.} 
We write these incentive constraints in matrix form as $Ux\ge 0$. Thus $U$ is an
$N\times M$ matrix, where $N=\sum_p |S_p|^2$.
The rows of $U$\fullver{, corresponding to the left-hand sides of the  constraints \eqref{eq:incentives},} are indexed by $(p,i,j)$, where $p$ is a player and $i,j\in S_p$ are a pair of $p$'s actions.
Denote by 
$U_s$ the column of $U$ corresponding to pure strategy profile $s$.
These incentive constraints, together with the constraints
\fullver{\begin{equation}\label{eq:prob}}\shortver{$}
x\ge 0, \; \sum_{s\in S} x_s =1\fullver{,\end{equation}}\shortver{$,}
which ensure that $x$ is a probability distribution, form a linear feasibility program that defines the set of CE.
}
\thesis{Correlated equilibrium (CE) is defined in Definition \ref{def:CE}.}
The problem of computing a maximum social welfare CE can be formulated as the LP
\begin{align}\label{primal}
\max\: & w^T x\tag{$P$}\\
Ux\thesis{&}\geq 0 \thesis{\nonumber\\}\nothesis{, \;}
x &\geq 0\thesis{\nonumber\\}\nothesis{, \;}
\sum_{s\in S} x_s=1\nonumber
\end{align}

\fullver{
Another objective of interest is the max-min welfare CE problem: computing a CE that maximizes the utility of the worst-off player.
\begin{align}\label{mmprimal}
\max \:r\\
\sum_s x_s u^p_s\geq r\qquad \forall p\\
Ux\geq 0 \thesis{\nonumber\\}\nothesis{, \;}
x\geq 0\thesis{\nonumber\\}\nothesis{, \;}
\sum_{s\in S} x_s=1\nonumber
\end{align}
}

Another solution concept of interest is \emph{coarse correlated equilibrium} (CCE).
Whereas  CE requires that  each player has no profitable deviation even if she takes into account the  signal she receives from the intermediary,
CCE only requires that each player has no profitable \emph{unconditional deviation}.
\begin{definition}
A correlated distribution $x$ is a \emph{coarse correlated equilibrium} (CCE) if it satisfies the following incentive constraints:
for each player $p$ and each of his actions $j\in S_p$,
\fullver{\begin{equation}}\shortver{we have $}
\sum_{(i,s_{-p})\in S} [u^p_{is_{-p}} - u^p_{js_{-p}}] x_{is_{-p}} \geq 0\fullver{.\end{equation}}\shortver{$.}
\end{definition}
We write these incentive constraints in matrix form as $Cx\ge 0$. Thus $C$ is an $(\sum_p|S_p|)\times M$ matrix.
By definition, a CE is also a CCE.

The problem of computing a maximum social welfare CCE can be formulated as the LP
\begin{align}\label{cprimal}
\max\: & w^T x\tag{$CP$}\\
Cx \thesis{&}\geq 0\thesis{\nonumber\\}\nothesis{, \;}
x &\geq 0\thesis{\nonumber\\}\nothesis{, \;}
\sum_{s\in S} x_s=1.\nonumber
\end{align}
\section{The Deviation-Adjusted Social Welfare Problem}

Consider the dual of \eqref{primal},
\begin{align}\label{dual}
\min\: &t \tag{$D$}\\
U^T y + w &\leq t\One\nonumber\\
y&\geq 0.\nonumber
\end{align}
We label the $(p,i,j)$-th element of $y\in \Real^N$ (corresponding to row $(p,i,j)$ of $U$) as $y^p_{i,j}$.
This is an LP with a polynomial number of variables and an exponential number of constraints. Given a separation oracle, we can solve it in polynomial time using the ellipsoid method.
A separation oracle needs to determine whether a given $(y,t)$ is feasible, and if not output a hyperplane that separates $(y,t)$ from the feasible set.
We focus on a restricted form
of separation oracles, which outputs a violated constraint for infeasible points.\footnote{This is a restriction because in general there exist separating hyperplanes
other than the violated constraints. For example \thesis{as we saw in Chapter \ref{ch:CE}, }\emcite{PR08JACM}'s algorithm for computing a sample CE uses a separation oracle that outputs a convex combination of the constraints as a separating hyperplane.}
Such a separation oracle needs to solve the
following problem: \begin{problem}\label{prob:optCE_sep}
Given $(y,t)$ with $y\geq 0$, determine if there exists an $s$ such that $(U_s)^T y+w_s > t$; if so output such an $s$.
\end{problem}
The left-hand-side expression $(U_s)^T y+w_s$ is the social welfare at $s$ plus the term $(U_s)^T y$.
Observe that the $(p,i,j)$-th entry of $U_s$ is $ u^p_s- u^p_{js_{-p}}$  if $s_p= i$ and is zero otherwise.
Thus $(U_s)^T y=\sum_p\sum_{j\in S_p} y^p_{s_p,j}\left(u^p_s- u^p_{js_{-p}}\right)$.
We now reexpress $(U_s)^T y+w_s$  in terms of \emph{deviation-adjusted utilities} and \emph{deviation-adjusted social welfare}.
\begin{definition}\label{def:adj_u_sw}
Given a game, and a vector $y\in\Real^{N}$
such that $y\geq 0$, the \emph{deviation-adjusted utility} for player $p$ under pure profile $s$ is
\[
\hat{u}^p_s(y)= u^p_s +\sum_{j\in S_p} y^p_{s_p,j}\left(u^p_s- u^p_{js_{-p}}\right).
\]
The deviation-adjusted social welfare is $\hat{w}_s(y)=\sum_p \hat{u}^p_s(y)$.
\end{definition}
By construction,
 the deviation-adjusted social welfare $\hat{w}_s(y)=\sum_pu^p_s+\linebreak\sum_p\sum_{j\in S_p}y^p_{s_p,j}\left(u^p_s- u^p_{js_{-p}}\right)= (U_s)^T y+w_s$.
Therefore, Problem \ref{prob:optCE_sep} is equivalent to the following \emph{deviation-adjusted social welfare problem}.
\begin{definition}
For a game representation, the \emph{deviation-adjusted social welfare problem} is the following: given an instance of the representation and rational vector $(y,t)\in\Rational^{N+1}$
such that $y\geq 0$, determine if there exists an $s$ such that the deviation-adjusted social welfare $\hat{w}_s(y)>t$; if so output such an $s$.
\end{definition}

\begin{proposition}
If the deviation-adjusted social welfare problem can be solved in polynomial time for a game representation, then so can the problem of computing the maximum social welfare CE.
\end{proposition}
\fullver{
\begin{proof}
Recall that an algorithm for Problem \ref{prob:optCE_sep} can be used as a separation oracle for \eqref{dual}.
Then we can apply the ellipsoid method  using the given algorithm for the deviation-adjusted social welfare problem as a separation oracle. This  solves \eqref{dual} in polynomial time.
By LP duality, the optimal objective of \eqref{dual} is the social welfare of the optimal CE.
The cutting planes generated during the ellipsoid method can then be used to compute such a CE with polynomial-sized support.
\nothesis{\qed}
\end{proof}
}

\thesis{We observe that our approach has certain similarities to the Ellipsoid Against Hope algorithm and its variants discussed in \autoref{ch:CE}:
both approaches are black-box approaches based on LP duality formulations of the respective problems, and both make use of the ellipsoid method to overcome
the exponential size of the LPs.
On the other hand, due to the different LP formulations of the sample CE problem and the optimal CE problem respectively, the two approaches require different separation oracles, which leads to the different requirements on the subroutines provided by the representation.
}

Let us consider interpretations of the dual variables $y$ and the deviation-adjusted social welfare of a game.
\hide{
\cite{hart1989existence} proved the existence of CE in any finite game by analyzing the LP formulation of CE and interpreting it as a two-player game between a primal player and a dual player.
Can we interpret a pair of primal and dual LPs like
\eqref{primal} and \eqref{dual}
 as a constant-sum game played by a player controlling the primal variables $x$ and a player controlling the dual variables $y$?
 One obstacle is that in \eqref{dual} the domain of $y$ is unbounded, so we cannot easily interpret it as a mixed strategy.
 Relevant is \cite{nau1990coherent} which gave another proof of the existence of CE. Their dual formulation has a unbounded domain for $y$,
 and they interpret such a $y$ as an arbitrage plan against the group of players in the game.
On the other hand, our LP formulation is different in that we have the social welfare as the objective of \eqref{primal}.

We now give one interpretation of the dual problem.}
The dual \eqref{dual} can be rewritten as
$\min_{y\geq 0}\max_s \tilde{w}_s(y)$. By weak duality, for a given $y\geq 0$ the maximum deviation-adjusted social welfare $\max_s \tilde{w}_s(y)$ is an upper bound on the
maximum social welfare CE. So the task of the dual \eqref{dual} is to find $y$ such that the resulting maximum deviation-adjusted social welfare gives the tightest bound.\footnote{An equivalent perspective is to view $y$ as Lagrange
multipliers, and the optimal deviation-adjusted SW problem as the Lagrangian
relaxation of \eqref{primal} given the multipliers $y$.
}
At optimum,
 $y$ corresponds to the concept of ``shadow prices'' from optimization theory; that is,
$y^p_{ij}$ equals the rate of change in the social welfare objective when the constraint $(p,i,j)$ is relaxed infinitesimally.
Compared to the maximum social welfare CE problem, the maximum deviation-adjusted social welfare problem
replaces the incentive constraints with a set of additional penalties  or rewards.
Specifically,
we can interpret $y$ as a set of nonnegative prices, one for each incentive constraint $(p,i,j)$ of \eqref{primal}.
At strategy profile $s$, for each incentive constraint $(p,i,j)$ we impose a penalty equal to $y^p_{ij}$ times the amount the constraint $(p,i,j)$ is violated
by $s$. Note that the penalty can be negative, and is zero if $s_p\neq i$.
Then $\tilde{w}_s(y)$ is equal to the social welfare of the modified game.

\begin{bf}Practical computation.\end{bf}
\thesis{We have seen from Chapters \ref{ch:Survey}, \ref{ch:AGG} and \ref{ch:CE} that t}\nothesis{T}he problem of computing the expected utility \nothesis{(EU)} given a mixed strategy profile has been established as an important subproblem for both the sample NASH problem and the sample CE problem, both in theory \nothesis{\cite{Daskalakis06,PR08JACM}} and in practice\nothesis{ \cite{BlumSheltonKoller,AGG-full}}.
Our results \thesis{in this chapter} suggest that the deviation-adjusted social welfare problem is of similar importance to the optimal CE problem.
This connection is more than theoretical: our algorithmic approach can be turned into a practical method for computing optimal CE.
In particular, although it makes use of the ellipsoid method, we can easily substitute a more practical method, such as simplex with column generation.
In contrast, \emcite{PR08JACM}'s algorithmic approach for reduced forms makes two nested applications of the ellipsoid method, and is less likely to be practical.
\fullver{Furthermore, even for representations without a polynomial-time algorithm for the deviation-adjusted social welfare problem, a promising direction would be to  formulate the deviation-adjusted social welfare problem as
a integer program or constraint program and solve using e.g. CPLEX.
}
\fullver{
\subsection{The Weighted Deviation-Adjusted Social Welfare Problem}
For the max-min welfare CE problem, we can form the dual of \eqref{mmprimal},
\begin{align}
\min \: t\label{mmdual}\\
U^Ty + \sum_p v_p u^p \leq t\One\label{eq:maxmin_dual_constr}\\
y\geq 0, \;
v\geq 0\fullver{\nonumber\\}\shortver{, \;}
\sum_p v_p=1.\notag
\end{align}
This is again an LP with polynomial number of variables and exponential number of constraints; specifically, block \eqref{eq:maxmin_dual_constr} 
is exponential. We observe that \eqref{eq:maxmin_dual_constr} is similar to the corresponding block in \eqref{dual}, except for the weighted sum
$\sum_p v_p u^p $ instead of the social welfare $w$.
Thus, in order to express the left-hand side of \eqref{eq:maxmin_dual_constr} we need notions slightly different from those given in Definition \ref{def:adj_u_sw}, which we call \emph{weighted deviation-adjusted utility} and \emph{weighted deviation-adjusted social welfare}.
\begin{definition}
Given a game, a vector $y\in\Real^{N}$
such that $y\geq 0$, and a vector $ v\in \Real^n$ such that $v\geq 0$ and $\sum_p v_p=1$,
the \emph{weighted deviation-adjusted utility} for player $p$ under pure profile $s$ is
\[\hat{u}^p_s(y,v)=v_p u^p_s +\sum_{j\in S_p} y^p_{s_p,j}(u^p_s- u^p_{js_{-p}}).
\]
The weighted deviation-adjusted social welfare is $\hat{w}_s(y,v)=\sum_p \hat{u}^p_s(y,v)$.
\end{definition}
Following analysis similar to that given above, the following problem serves as a separation oracle of LP \eqref{mmdual}.
\begin{definition}
For a game representation,
the \emph{weighted deviation-adjusted social welfare problem} is the following: given an instance of the representation, and rational vector $(y,v,t)\in\Rational^{N+n+1}$
such that $y\geq 0$, $v\geq 0$ and $\sum_p v_p=1$, determine if there exists an $s$ such that the deviation-adjusted social welfare $\hat{w}_s(y)>t$; if so output such an $s$.
\end{definition}
\begin{proposition}
If the weighted deviation-adjusted social welfare problem can be solved in polynomial time for a game representation, then the problem of computing the max-min welfare CE
is in polynomial time for this representation.
\end{proposition}
It is straightforward to see that the deviation-adjusted social welfare problem reduces to the weighted deviation-adjusted social welfare problem.
In all representations that we consider in this chapter, the weighted and unweighted versions have the same structure and thus the same complexity.
}
\subsection{The Coarse Deviation-Adjusted Social Welfare Problem}
For the optimal social welfare CCE problem, we can
form the dual of \eqref{cprimal}
\begin{align}\label{cdual}
\min\:  &t\\
C^T y + w &\leq t\One\nonumber\\
y&\geq 0\nonumber
\end{align}
\begin{definition}
We label the $(p,j)$-th element of $y$ as $y^p_j$. 
Given a game, and a 
vector $y\in\Real^{\sum_p|S_p|}$ such that $y\geq 0$,
the \emph{coarse deviation-adjusted utility} for player $p$ under pure profile $s$ is
\fullver{\[}\shortver{$}
\tilde{u}^p_s(y)= u^p_s +\sum_{j\in S_p} y^p_{j}(u^p_s- u^p_{js_{-p}})\fullver{.\]}\shortver{$.}
The coarse deviation-adjusted social welfare is $\tilde{w}_s(y)=\sum_p \tilde{u}^p_s(y)$.
\end{definition}
\begin{proposition}
If the coarse deviation-adjusted social welfare problem can be solved in polynomial time for a game representation, then the problem of computing the maximum social welfare CCE
is in polynomial time for this representation.
\end{proposition}

The coarse deviation-adjusted social welfare problem reduces to the deviation-adjusted social welfare problem.
To see this, given an input 
vector $y$ for the coarse deviation-adjusted social welfare problem, we can construct an input vector $y'\in \Rational^N$ for the deviation-adjusted social welfare problem with
$y'^p_{ij} = y_j^p$ for all $p\in \mathcal{N}$ and  $i,j\in S_p$.
\section{The Deviation-Adjusted Social Welfare Problem for Specific Representations}


In this section we study the deviation-adjusted social welfare problem and its variants on specific representations.
Depending on the representation, the deviation-adjusted social welfare problem is not always solvable in polynomial time. Indeed, \emcite{PR08JACM} showed that for many
representations the problem of optimal CE is NP-hard.
Nevertheless, for such representations we can often identify tractable subclasses of games.
We will argue that the deviation-adjusted social welfare problem is a more useful formulation for identifying tractable classes of games
than the separation problem formulation of \emcite{PR08JACM}, as the latter only applies to reduced-form-based representations.

\subsection{Reduced Forms}\label{sec:reduced_form}

\emcite{PR08JACM}   gave the following reduced form characterization of representations.
\begin{definition}[\cite{PR08JACM}]
Consider a game $G=(\mathcal{N}$, $\{S_p\}_{p\in \mathcal{N}}, \{u^p\}_{p\in \mathcal{N}})$. For $p=1,\ldots ,n$, let $P_p=\{C_p^1\ldots C_p^{r_p}\}$ be a partition of $S_{-p}$ into $r_p$ classes. The set $\mathcal{P}=\{P_1,\ldots,P_n\}$ of partitions is a \emph{reduced form} of $G$ if $u^p_s=u^p_{s'}$ whenever (1) $s_p=s'_p$ and (2) both $s_{-p}$ and $s'_{-p}$ belong to the same class in $P_p$. The \emph{size} of a reduced form is the number of classes in the partitions plus the bits required to specify a payoff value for each tuple $(p, k,\ell)$
where $1\leq p\leq n$, $1\leq k\leq r_p$ and $\ell\in S_p$.
\end{definition}
Intuitively, the reduced form imposes the condition that $p$'s utility for choosing an action $s_p$ depends only on which \emph{class} in the partition $P_p$ the profile
of the others' actions belongs to.

\emcite{PR08JACM} showed that several compact representations such as graphical games and anonymous games have natural reduced forms
whose sizes are (roughly) equal to the sizes of the representation.
We say such a compact representation has a \emph{concise reduced form}.
Intuitively, such a reduced form describes the structure of the game's utility functions.
\fullver{
\begin{example}
\thesis{Recall from \autoref{sec:lit_static} that a}\nothesis{A} graphical game \cite{graphical} is associated with a graph $(\mathcal{N},E)$, such that player $p$'s utility depends only on her action
and the actions of her neighbors in the graph. The sizes of the utility functions are exponential only in the degrees of the graph.
Such a game has a natural reduced form where the classes in $P_p$ are identified with the pure profiles of $p$'s neighbors, i.e., $s_{-p}$
and $s'_{-p}$ belong to the same class if and only if they agree on the actions of $p$'s neighbors.
The size of the reduced form is exactly the number of utility values required to specify the graphical game's utility functions.\qed
\end{example}
}

Let $\mathcal{S}_p(k,\ell)$ denote the set of pure strategy profiles $s$ such that $s_p=\ell$ and $s_{-p}$ is in the $k$-th class $C_p^k$ of $P_p$,
and let $u^p_{(k,\ell)}$ denote the utility of $p$ for that set of strategy profiles.
\emcite{PR08JACM} defined the following \emph{Separation Problem} for a reduced form.
\begin{definition}[\cite{PR08JACM}]
Let $\mathcal{P}$ be a reduced form for game $G$. The \emph{Separation Problem} for $\mathcal{P}$ is the following:
Given rational numbers $\gamma_p(k,\ell)$ for all $p\in \{1, \ldots, n\}$, $k\in \{1, \ldots, r_p\}$, and $\ell\in S_p$, is there a pure strategy profile $s$ such that
$
\sum_{p,k,\ell: s\in \mathcal{S}_p(k,\ell)} \gamma_p(k,\ell)<0 ?
$
If so, find such\fullver{ an} $s$.
\end{definition}
Since  $s\in \mathcal{S}_p(k,\ell)$ implies $s_p=\ell$, the left-hand side of the  above expression is equivalent to $\sum_p\sum_{k:s\in \mathcal{S}_p(k,s_p)} \gamma_p(k,s_p)$.
Furthermore, since $s$ belongs to exactly one class in $P_p$, the expression is a sum of exactly $n$ summands\fullver{, one for each player}.

\emcite{PR08JACM} proved that if the separation problem can be solved in polynomial time, then a CE that maximizes a given linear objective in the players' utilities
can be computed in time polynomial in the size of the reduced form.
How does \emcite{PR08JACM}'s sufficient condition relate to ours, provided that the game has a concise reduced form? We show that
the class of reduced form games for which our \fullver{weighted} deviation-adjusted  social welfare  problem is polynomial-time solvable contains the class for which the separation problem is polynomial-time solvable.

\begin{proposition}\label{prop:wdsw_reduced_form}
Let $\mathcal{P}$ be a reduced form for game $G$. Suppose the separation problem can be solved in polynomial time. Then
the \fullver{weighted} deviation-adjusted  social welfare problem can be solved in time polynomial in the size of the reduced form.
\end{proposition}
\fullver{
\begin{proof}
First we observe that if a game $G$ has a reduced form $\mathcal{P}$, then its deviation-adjusted utilities \fullver{(and weighted deviation-adjusted utilities)} also satisfy the partition structure
specified by $\mathcal{P}$, i.e., given $y$ and $v$, the weighted deviation-adjusted utility $\hat{u}^p_s(y,v)$ depends only on a player's action $s_p$ and the class in $P_p$ that $s_{-p}$ belongs to.
To see why, suppose $s_{-p}\in C_p^k$. Then
\begin{align*}
\hat{u}^p_{\ell s_{-p}}(y,v)&=v_p u^p_{\ell s_{-p}} +\sum_{j\in S_p} y^p_{\ell,j}(u^p_{\ell s_{-p}}- u^p_{js_{-p}})\\
&= v_p u^p_{(k,\ell)} + \sum_{j\in S_p} y^p_{\ell,j} (u^p_{(k,\ell)} -u^p_{(k,j)}),
\end{align*}
which depends only on $\ell$ and $k$.
This proves the following, which will be useful later.
\begin{lemma}\label{lem:wdu_reduced_form}
Let $\mathcal{P}$ be a reduced form for game $G$.
\begin{enumerate}
\item For all $y\in \Real^N$, $v\in \Real^n$, for all players $p$, $s_p\in S_p$, and for all $s_{-p}, s'_{-p}\in S_{-p}$, if
$s_{-p}$ and $s'_{-p}$ are in the same class in $P_p$
then the weighted deviation-adjusted utilities
$\hat{u}^p_{s_p,s_{-p}}(y,v)=\hat{u}^p_{s_p,s'_{-p}}(y,v)$.

\item Write the weighted deviation-adjusted utility for player $p$, given her pure strategy $\ell\in S_p$ and class $C_p^k$,
as $\hat{u}^p_{(k,\ell)}(y,v)$ (well defined by the above). We have
\[
\hat{u}^p_{(k,\ell)}(y,v) \equiv v_p u^p_{(k,\ell)} + \sum_{j\in S_p} y^p_{\ell,j} (u^p_{(k,\ell)} -u^p_{(k,j)}).
\]
\end{enumerate}
\end{lemma}

Given an instance of the weighted deviation-adjusted  social welfare problem with a game with reduced form $\mathcal{P}$ and rational vectors $y\in \Real^N$, $v\in \Real^n$ and $t\in \Real$, we construct an instance of the separation problem by letting $\gamma_p(k,\ell) = t/n - \hat{u}^p_{(k,\ell)}(y,v)$,
where $\hat{u}^p_{(k,\ell)}(y,v)$ is as defined in Lemma \ref{lem:wdu_reduced_form} and can be efficiently computed given the reduced form.
 Recall that the separation problem asks for pure profile $s$ such that $\sum_{p,k,\ell: s\in \mathcal{S}_p(k,\ell)} \gamma_p(k,\ell)<0$, the left hand side of which
 is a sum of $n$ terms.
 By construction, for all  $s$,
 $\sum_{p,k,\ell: s\in \mathcal{S}_p(k,\ell)} \gamma_p(k,\ell)<0 $ if and only if
  $\sum_p\sum_{k:s\in \mathcal{S}_p(k,s_p)} \left(t/n - \hat{u}^p_{(k,s_p)}(y,v)\right)<0$, and since the left hand side  is a sum of $n$ terms,
   this holds if and only if
  $\hat{w}^p_s(y,v) >t$.
Therefore  the weighted deviation-adjusted social welfare problem instance  has a solution $s$ if and only if the
corresponding separation problem instance has a solution $s$, and a polynomial-time algorithm for the separation problem can be used to
solve
the weighted deviation-adjusted social welfare problem in polynomial time.
\nothesis{\qed}
\end{proof}
}

We now compare the the \fullver{weighted} deviation-adjusted social welfare problem with the optimal social welfare problem for these representations.
We observe \fullver{from Lemma \ref{lem:wdu_reduced_form}} that the \fullver{weighted} deviation-adjusted social welfare problem can be formulated as an instance of the optimal social welfare problem on another game with the same reduced form
but different payoffs. Can we claim that the existence of a polynomial-time algorithm for the optimal social welfare problem for a representation implies the
existence of a polynomial-time algorithm for the \fullver{weighted} social welfare problem (and thus the optimal CE problem)?
This is not necessarily the case, because the representation might impose certain structure on the utility functions that are not captured by the reduced forms, and
the polynomial-time algorithm for the optimal social welfare problem could depend on the existence of such structure. The \fullver{weighted} deviation-adjusted social welfare problem might no longer exhibit such structure and thus might not be solvable using the given algorithm.

Nevertheless, if we consider a game representation that is ``completely characterized''
by its reduced forms,  the \fullver{weighted} deviation-adjusted social welfare problem is equivalent to the decision version of the optimal social welfare outcome problem for that representation.
To make this more precise, we say a game representation is a \emph{reduced-form-based representation} if
there exists a mapping from instances of the representation to reduced forms such that it maps each instance to a concise reduced form of that instance,
and if we take such a reduced form and change its payoff values arbitrarily, the resulting reduced form
is a concise reduced form of  another instance of the representation.
\begin{corollary}\label{coro:reduced_form_corrspondence}
For a reduced-form-based representation, if there exists a polynomial-time algorithm for the optimal social welfare problem, then the optimal
social welfare CE problem and the max-min welfare CE problem can be solved in polynomial time.
\end{corollary}
Of course, this can be derived using the separation problem for reduced forms without the deviation-adjusted social welfare formulation.
On the other hand, the deviation-adjusted social welfare formulation can be applied to representations without concise reduced forms.
In fact, we will use it to show below that the connection between the optimal social welfare problem and the optimal CE problem
applies to a wider classes of representations
than just reduced-form-based representations.

\subsection{Linear Reduced Forms}
One class of representations that does not have concise reduced forms are those that represent utility functions as sums of other functions, such as polymatrix games and the hypergraph games of \emcite{PR08JACM}.
In this section we characterize these representations using linear reduced forms, showing that linear-reduced-form-based representations satisfy a property similar
to Corollary \ref{coro:reduced_form_corrspondence}.

Roughly speaking, a linear reduced form has multiple partitions for each agent, rather than just one; an agent's overall utility is a sum over utility functions defined on each of that agent's partitions.
\begin{definition}
Consider a game $G=(\mathcal{N}, \{S_p\}_{p\in \mathcal{N}}, \{u^p\}_{p\in \mathcal{N}})$. For $p=1,\ldots ,n$, let $P_p=\{P_{p,1},\ldots,P_{p,t_p}\}$, where
$P_{p,q}=\{C_{p,q}^1\ldots C_{p,q}^{r_{pq}}\}$ is a partition of $S_{-p}$ into $r_{pq}$ classes. The set $\mathcal{P}=\{P_1,\ldots,P_n\}$ is a \emph{linear reduced form} of $G$ if for each $p$ there exist $u^{p,1},\ldots, u^{p,t_p}\in \Real^M$ such that for all $s$, $u^p_s=\sum_q u^{p,q}_s$,
and for each $q\leq t_p$, $u^{p,q}_s=u^{p,q}_{s'}$ whenever (1) $s_p=s'_p$ and (2) both $s_{-p}$ and $s'_{-p}$ belong to the same class in $P_{p,q}$. The \emph{size} of a reduced form is the number of classes in the partitions plus the bits required to specify a number for each tuple
$(p,q,k,\ell)$ where $1\leq p\leq n$, $1\leq q\leq t_p$, $1\leq k\leq r_{pq}$ and  $\ell\in S_p$.
\end{definition}
We write $u^{p,q}_{(k,\ell)}$ for the value corresponding to tuple $(p,q,k,\ell)$, and for  $\mathbf{k}=(k_1,\ldots,k_{t_p} )$
we write $u^{p}_{(\mathbf{k},\ell)} \equiv \sum_q u^{p,q}_{(k_q,\ell)}$.

\begin{example}[polymatrix games]\label{ex:polymatrix_lrf}
\thesis{Recall from \autoref{sec:lit_static} that in}\nothesis{In} a polymatrix game, each player's utility is the sum of utilities resulting from her bilateral interactions with each of the $n-1$ other players:
$
u^p_s = \sum_{p'\neq p} e_{s_p}^T A^{pp'} e_{s_{p'}}
$
where  $A^{pp'}\in \Real^{|S_p|\times| S_{p'}|}$ and $ e_{s_p}\in \Real^{|S_p|}$ is the unit vector corresponding to $s_p$.
The utility functions of such a representation require only $\sum_{p,p'\in \mathcal{N}}|S_p|\times| S_{p'}|$ values to specify.
Polymatrix games do not have a concise reduced-form encoding, but can easily be written as linear-reduced-form games. Essentially, we create one partition for every matrix game that an agent plays, with each class differing in the action played by the other agent who participates in that matrix game, and containing all the strategy profiles that can be adopted by all of the other players.
Formally, given a polymatrix game, we construct its linear reduced form
with $P_p=\{P_{p,q}\}_{q\in \mathcal{N}\setminus\{ p\}}$, and $P_{p,q} =\{ C_{p,q}^\ell\}_{\ell\in S_q}$ with $C_{p,q}^\ell=\{s_{-p}|s_q=\ell\}$.\qed
\end{example}

Most of the results in Section \ref{sec:reduced_form} straightforwardly translate to linear reduced forms.

\fullver{
\begin{lemma}\label{lem:wdu_linear_reduced_form}
Let $\mathcal{P}$ be a linear reduced form for game $G$.
Then for all $y\in \Real^N$, $v\in \Real^n$, for all players $p$, there exist $\hat{u}^{p,1}(y,v),\ldots, \hat{u}^{p,t_p}(y,v)\in \Real^M$ such that the weighted deviation-adjusted utilities $\hat{u}^p(y,v)=\sum_q \hat{u}^{p,q}(y,v)$, and
for all $q\leq t_p$,
$s_p\in S_p$ and $s_{-p}, s'_{-p}\in S_{-p}$, if
$s_{-p}$ and $s'_{-p}$ are in the same class in $P_{p,q}$,
then
$\hat{u}^{p,q}_{s_p,s_{-p}}(y,v)=\hat{u}^{p,q}_{s_p,s'_{-p}}(y,v)$.

Write the weighted deviation-adjusted utility for player $p$, her pure strategy $\ell\in S_p$ and classes $C_{p,1}^{k_1}, \ldots, C_{p,t_p}^{k_{t_p}}$
as $\hat{u}^p_{(\mathbf{k},\ell)}(y,v)$ where $\mathbf{k}=(k_1,\ldots,k_{t_p} )$. Furthermore, we have
\[
\hat{u}^p_{(\mathbf{k},\ell)}(y,v) \equiv v_p u^p_{(\mathbf{k},\ell)} + \sum_{j\in S_p} y^p_{\ell,j} (u^p_{(\mathbf{k},\ell)} -u^p_{(\mathbf{k},j)}).
\]
\end{lemma}
}

\begin{corollary}\label{coro:linear_reduced_form_corrspondence}
For a linear-reduced-form-based representation, if there exists a polynomial-time algorithm for the optimal social welfare problem, then the optimal
social welfare CE problem and the max-min welfare CE problem can be solved in polynomial time.
\end{corollary}

\subsubsection{Graphical Polymatrix Games}
A polymatrix game may have graphical-game-like structure: player $p$'s utility may depend only on a subset of the other player's actions.
In terms of utility functions, this corresponds to $A^{pp'}=0$ for certain pairs of players $p,p'$.
As with graphical games, we can construct the (undirected) graph $G=(\mathcal{N},E)$ where there is an edge $\{p,p'\}\in E$ if $A^{pp'}\neq 0$ or$A^{p'p}\neq 0$.
We call such a game a graphical polymatrix game.
This can also be understood as a graphical game where each player $p$'s utility is the sum of bilateral interactions with her neighbors.

A tree polymatrix game is a graphical polymatrix game whose corresponding graph is a tree. Consider the optimal CE problem on tree polymatrix games.
Since such a game is also a tree graphical game,
\emcite{PR08JACM}'s optimal CE algorithm for tree graphical games can be applied. However, this algorithm does not run in polynomial time, because
the representation size of tree polymatrix games can be exponentially smaller than that of the corresponding graphical game (which grows exponentially in the degree of the graph).
\fullver{However, we can give a different polynomial-time algorithm for this problem.}\shortver{Nevertheless, we give an polynomial-time algorithm for the deviation-adjusted social welfare problem for such games, which then implies the following theorem.}
\begin{theorem}
Optimal CE in tree polymatrix games can be computed in polynomial time.
\end{theorem}
\fullver{
\begin{proof}
It is sufficient to give an algorithm for the deviation-adjusted social welfare problem.
Using an argument similar to that given in Example \ref{ex:polymatrix_lrf},
tree polymatrix games have a natural linear reduced form, and it is straightforward to verify that tree polymatrix games
are a linear-reduced-form-based representation.
By Corollary \ref{coro:linear_reduced_form_corrspondence} it is sufficient to construct an algorithm for the optimal social welfare problem.

Let $N_p$ be the set of players in the subtree rooted at $p$.
 Suppose  $p$'s parent in the tree is $q$. Let the  \emph{social welfare contribution} of $N_p$ be the social welfare of players in $N_p$ minus $e_{s_p}^T A^{pq} e_{s_{q}}$. Let the social welfare contribution of the root player be the social welfare of $\mathcal{N}$. Then the social welfare contribution of $N_p$ depends solely on the pure strategy profile restricted to $N_p$.

The following dynamic programming algorithm solves the optimal social welfare problem in polynomial time. We go
from the leaves to the root of the tree.
 Each child $q $ of $p$ passes to its parent the message $\{w^{N_{q},s_q}\}_{s_q\in S_q}$, where $w^{N_{q},s_q}$ is the optimal social welfare contribution of $N_{q}$ provided that $q$ plays $s_q$.
 Given the messages from all of $p's$ children $q_1,\ldots, q_k$,
 we can compute the message of $p$ as follows: for each $s_p\in S_p$,
 \begin{align*}
 w^{N_{p},s_p}& = \max_{s_{q_1},\ldots,s_{q_k}}\sum_{j=1}^k \left[w^{N_{q_j},s_{q_j}} + e_{s_p}^TA^{p,{q_j}} e_{s_{q_j}}\right] \\
 &=\sum_{j=1}^k \max_{s_{q_j}}\left[w^{N_{q_j},s_{q_j}} + e_{s_p}^TA^{p,{q_j}} e_{s_{q_j}}\right].
\end{align*}
The second equality is due to the fact that the $j$-th summand depends only on $s_{q_j}$.
It is straightforward to verify that the optimal social welfare is
$\max_{s_r} w^{N_r,s_r}$ where $r$ is the root player, and that the algorithm runs in polynomial time.
The corresponding optimal pure strategy profile can be constructed by going from the root to the leaves. \nothesis{\qed}
\end{proof}
This algorithm can be straightforwardly extended to yield a polynomial-time algorithm for optimal CE in  graphical polymatrix games with constant treewidth,
for hypergraphical games \cite{PR08JACM} on acyclic hypergraphs, and more generally for hypergraphs with constant hypertree-width.
}
\subsection{Representations with Action-Specific Structure}\label{sec:optCE_act_spec}
The above results for reduced forms and linear reduced forms crucially depend on the fact that the partitions (i.e., the structure of the utility functions) depend on $p$ but
do not depend on the action chosen by player $p$.

There are representations whose utility functions have action-dependent structure, including
congestion games \cite{congestion}, local effect games\\
 \cite{localeffect}, and action-graph games \cite{AGG-full}.
For such representations, we can define a variant of the reduced form that has action-dependent partitions.
\thesis{For example:
\begin{definition}
Consider a game $G=(\mathcal{N}, \{S_p\}_{p\in \mathcal{N}}, \{u^p\}_{p\in \mathcal{N}})$. For $p=1,\ldots ,n$, $\ell\in S_p$, let $P_{p,\ell}=\{P_{p,\ell,1},\ldots,P_{p,\ell,t_{p\ell}}\}$, where
$P_{p,\ell,q}=\{C_{p,\ell,q}^1\ldots C_{p,\ell,q}^{r_{p\ell q}}\}$ is a partition of $S_{-p}$ into $r_{p\ell q}$ classes. The set $\mathcal{P}=\{P_{p,\ell}\}_{p\in \mathcal{N}, \ell\in S_p}$ is a \emph{action-specific linear reduced form} of $G$ if for each $p,\ell$ there exist $u^{p,\ell,1},\ldots, u^{p,\ell,t_{p\ell}}\in \Real^M$ such that for each $p\in \mathcal{N}$, $\ell\in S_p$, and $q\leq t_p$,
\begin{enumerate}
\item for all $s_{-p}\in S_{-p}$, $u^p_{\ell s_{-p}}=\sum_q u^{p,\ell,q}_{\ell s_{-p}}$;

\item $u^{p,\ell,q}_{\ell s_{-p}}=u^{p,\ell,q}_{\ell s'_{-p}}$ whenever
 both $s_{-p}$ and $s'_{-p}$ belong to the same class in $P_{p,\ell,q}$.

\end{enumerate}
The \emph{size} of a reduced form is the number of classes in the partitions plus the bits required to specify a number for each tuple
$(p,q,k,\ell)$ where $1\leq p\leq n$, $1\leq q\leq t_{p\ell}$, $1\leq k\leq r_{p\ell q}$ and  $\ell\in S_p$.
\end{definition}
}
However, unlike both the reduced form and linear reduced form,  the \fullver{weighted} deviation-adjusted utilities no longer satisfy the same partition structure as the utilities.
Intuitively, the \fullver{weighted} deviation-adjusted utility at $s$ has contributions from the utilities of the strategy profiles when player $p$ deviates to different actions.
Whereas for linear reduced forms these deviated strategy profiles correspond to the same class as $s$ in the partition,
we now consider different partitions for each action to which $p$ deviates.
As a result the \fullver{weighted} deviation-adjusted social welfare problem has a more complex form that the optimal social welfare problem.

\subsubsection{Singleton Congestion Games}
\thesis{As mentioned in Chapters \ref{ch:Survey} and \ref{ch:pure}, }\emcite{ieong2005fac} studies a class of games called singleton congestion games and showed that the optimal PSNE can be computed in polynomial time.
Such a game can be formulated as an instance of congestion games where each action contains a single resource,
or an instance of symmetric AGGs where the only edges are self edges.

Formally, a singleton congestion game is specified by $( \mathcal{N}, \mathcal{A}, \{f^\alpha\}_{\alpha\in\mathcal{A} } )$ where $\mathcal{N}={1,\ldots,n}$ is the set of players,
$\mathcal{A}$ the set of actions, and for each action $\alpha\in\mathcal{A}$,  $f^\alpha: [n]\rightarrow \Real$.
The game is symmetric; each player's set of actions $S_p\equiv \mathcal{A}$. Each strategy profile $s$ induces an action count $c(\alpha)=|\{p|s_p=\alpha\}|$ on each $\alpha$: the number of players playing action $\alpha$. Then the utility of a player that chose $\alpha$ is
$f^\alpha(c(\alpha))$.
The representation requires  $O(|\mathcal{A}| n)$ numbers to specify.

We now show that the optimal social welfare CCE problem can be computed in polynomial time for singleton congestion games.
Before attacking the problem, we first note
 that the optimal social welfare problem can be solved in polynomial time by a relatively straightforward dynamic-programming algorithm which
 is a simplified version of \emcite{ieong2005fac}'s algorithm for optimal PSNE in singleton congestion games.
\fullver{First observe that  the social welfare of a strategy profile can be written in terms of the action counts:
\[w_s = \sum_\alpha c(\alpha) f^\alpha (c(\alpha)).\]
The optimal social welfare problem is equivalent to finding a vector of action counts that sums to $n$ and maximizes the above expression. The social welfare can be further
decomposed into contributions from each action $\alpha$. The dynamic-programming algorithm starts with a single action and adds one action at a time until all actions are added.
At each iteration, it maintains a set of tuples $\{(n', w^{n'}) \}_{1\leq n'\leq n}$, specifying that the best social welfare contribution from the current set of actions is $w^{n'}$ when exactly $n'$ players chose actions in the current set.

Consider the optimal social welfare CCE problem.
}
Can we leverage the algorithm for the optimal social welfare problem to solve the coarse deviation-adjusted social welfare problem?
Our task here is slightly more complicated: in general the coarse deviation-adjusted social welfare problem no longer has the same symmetric structure due to the fact that
$y$ can be asymmetric.
However, when $y$ is player-symmetric (that is, $y^p_j=y^{p'}_j$ for all pairs of players $(p,p')$), then we
recover symmetric structure.

\begin{lemma}\label{lem:optCE_scg_sep}
Given a singleton congestion game and player-symmetric input $y$, the coarse deviation-adjusted social welfare problem  can be solved in polynomial time.\end{lemma}
\fullver{
\begin{proof}
The coarse deviation-adjusted social welfare can be written as
\begin{align*}
\tilde{w}_s(y) &= \sum_p u^p_s(1+\sum_{j\neq s_p} y^p_j) - \sum_p\sum_{j\neq s_p} y^p_j u^p_{js_{-p}}\\
&=
\sum_{\alpha\in \mathcal{A}} \left[c(\alpha)f^\alpha(c(\alpha))\left(1+\sum_{j\neq \alpha} y^p_j\right)
-(n-c(\alpha))f^\alpha(c(\alpha)+1)y^p_\alpha\right].
\end{align*}
The contribution from each action $\alpha$ depends only on $c(\alpha)$. Therefore, using a similar dynamic-programming algorithm as above we
can solve the coarse deviation-adjusted social welfare problem in polynomial time. \nothesis{\qed}
\end{proof}
}

Therefore if we can guarantee that during a run of ellipsoid method for \eqref{cdual} all input queries $y$ to the separation oracle are symmetric,
then we can apply Lemma \ref{lem:optCE_scg_sep} to solve the problem in polynomial time.
We observe that for any symmetric game, there must exist a \emph{symmetric} CE that optimizes the social welfare. This is because given an optimal CE we can
create a mixture of permuted versions of this CE, which must itself be a CE by convexity, and must also achieve the same social welfare by symmetry.
However, this argument in itself does not guarantee that the $y$ we obtain by the method above will be symmetric.
Instead, we observe that if we solve \eqref{cdual} using a ellipsoid method with a player-symmetric initial ball, and
use a separation oracle that returns a player-symmetric cutting plane, then the query points $y$ will be player-symmetric.
We are able to construct such a separation oracle using a symmetrization argument.
\begin{theorem}\label{thm:optCE_scg}
Given a singleton congestion game, the optimal social welfare CCE can be computed in polynomial time.
\end{theorem}
\nothesis{\fullver{The proof is given in Appendix \ref{app:scg_proof}.}}
\thesis{\begin{proof}
As argued in Section \ref{sec:optCE_act_spec},
it is sufficient to construct a separation oracle for \eqref{cdual} that returns a player-symmetric cutting plane.
The cutting plane corresponding to a pure strategy profile solution $s$ of the coarse deviation-adjusted social welfare problem  is not player-symmetric in general; but we can symmetrize it by constructing a mixture of permutations of $s$. Since by symmetry each permuted version of $s$ correspond to a violated constraint, the resulting cutting plane is still correct and is symmetric. Enumerating all permutations over players would be exponential, but it turns out that for our purposes it is sufficient to use a small set of permutations.

Formally, let $\pi_i$ be the permutation over the set of players $\mathcal{N}$ that maps each $p$ to $p+ i \mod n$. Then the set  of permutations
$\{\pi_i\}_{0\leq i\leq n-1}$ corresponds to the cyclic group.

Suppose $s$ is a solution of the coarse deviation-adjusted social welfare problem with symmetric input $y$. The corresponding cut (violated constraint) is
$(C_s)^T y + w_s\leq t$. Recall that the $(p,j)$-th entry of $C_s$ is $C_s^{p,j}=(u^p_s-u^p_{js_{-p}})$.
For a permutation $\pi$ over $\mathcal{N}$, write $s^\pi$ the permuted profile induced by $\pi$, i.e. $s^\pi=(s_{\pi(1)},\ldots,s_{\pi(n)})$.
Then $s^\pi$ is also a solution of the coarse deviation-adjusted social welfare problem.
Form the following  convex combination of $n$ of the constraints of \eqref{cdual}:
\[
\frac{1}{n}\sum_{i=0}^{n-1} \left[(C_{s^{\pi_i}})^T y+ w_{s^{\pi_i}}\right] \leq t
\]
The left-hand side can be simplified to $w_s+(\overline{C}_s)^T y$ where $\overline{C}_s=\frac{1}{n}\sum_{i=0}^{n-1} C_{s^{\pi_i}}$.
We claim that this cutting plane is player-symmetric, meaning $\overline{C}_s^{p,j}=\overline{C}_s^{p',j}$ for all pairs of players $p,p'$ and all $j\in \mathcal{A}$.
This is because
\begin{align*}
\overline{C}_s^{p,j} & = \frac{1}{n}\sum_{i=0}^{n-1}C_{s^{\pi_i}}^{p,j}
= \frac{1}{n}\sum_{i=0}^{n-1} (u^p_{s^{\pi_i}}-u^p_{js^{\pi_i}_{-p}})\\
&= \frac{1}{n}\left[\sum_{\alpha\neq j} c(\alpha)f^\alpha(c(\alpha)) - (n-c(j))f^j(c(j)+1)\right]
=\overline{C}_s^{p',j}.
\end{align*}
This concludes the proof.
\nothesis{\qed}
\end{proof}
}
\fullver{Our approach for singleton congestion games crucially depends on the fact that the coarse deviation profile $y^p_j$ does not care which action  it is deviating from.
This allowed us to (in the proof of Lemma \ref{lem:optCE_scg_sep}) decompose the coarse deviation-adjusted social welfare into terms that only depend on the action count on one action.
The same approach cannot be directly applied to solve the optimal CE problem, because then the deviation profile would give a different
$y^p_{ij}$ for each action $i$ that $p$ deviates from, and the resulting expression for deviation-adjusted social welfare would involve summands that
depend on the action counts on pairs of actions.
}

\thesis{An interesting future direction is to explore whether our approach for singleton congestion games can be generalized to other classes of symmetric games, such as symmetric AGGs with bounded treewidth.}

\fullver{
\section{Conclusion and Open Problems}

We have proposed an algorithmic approach for solving the optimal correlated equilibrium problem in succinctly represented games, substantially extending a previous approach due to \emcite{PR08JACM}. In particular, we showed that the optimal CE problem is tractable when the \emph{deviation-adjusted social welfare problem} can be solved in polynomial time.
We generalized the reduced forms of \emcite{PR08JACM} to show that if a representation can be characterized by ``linear reduced forms'', i.e. player-specific linear functions over partitions, then for that representation, the deviation-adjusted social welfare problem can be reduced to the optimal social welfare problem. Leveraging this result, we showed that the optimal CE problem is tractable in graphical polymatrix games on tree graphs. We also considered the problem of computing the optimal \emph{coarse correlated equilibrium}, and derived a similar sufficient condition. We used this condition to prove that the optimal CCE problem is tractable for singleton congestion games.
}

\fullver{
Our work points the way to a variety of open problems, which we briefly summarize here.

\begin{bf}Price of Anarchy.\end{bf}
Our results imply that for compactly represented games with polynomial-time algorithms for the optimal social welfare problem and the weighted deviation-adjusted social welfare problem, the Price of Anarchy (POA) for correlated equilibria (i.e., the ratio of social welfare under the best outcome and the worst correlated equilibrium) can be computed in polynomial time.
Similarly for the Price of Total Anarchy (i.e., the ratio of social welfare under the best outcome and the worst coarse correlated equilibrium).
There is an extensive literature on proving bounds on the POA for various solution concepts and for various classes of games. One line of research that is particularly relevant to our work is the ``smoothness bounds'' method pioneered by 
\emcite{Roughgarden09POA}.
In particular, that work showed that if a certain smoothness relation can be shown to hold for a class of games, then it can be used to prove an upper bound on POA for these games that holds for many solution concepts including pure and mixed NE, CE and CCE.
More recently, \emcite{Nadav10pdPOA} gave a primal-dual LP formulation for proving POA bounds and showed that finding the best smoothness coefficients corresponds to
the dual of the LP for the POA for average coarse correlated equilibrium (ACCE), a weaker solution concept than CCE.
The primal-dual LP formulation of \emcite{Nadav10pdPOA} and our LPs \eqref{primal} and \eqref{dual} are equivalent up to scaling; however whereas \emcite{Nadav10pdPOA} focused on the task
of proving POA upper bounds for classes of games, here we focus on computing the optimal CE / CCE and POA for individual games.
One interesting direction is to use our algorithms together with an game instance generator to automatically find game instances with
large POA, thus improving the lower bounds on POA for given classes of games.
}

\fullver{
\begin{bf}Complexity separations.\end{bf}
We have shown that for singleton congestion games, the optimal social welfare problem and the optimal CCE problem are tractable while
the complexity of the optimal CE problem is unknown.  An open problem is to prove a separation of the complexities of these problems for singleton congestion games or for another class.
Another related problem is the optimal PSNE problem, which can be thought of as the optimal CE problem plus integer constraints on $x$.
We do not know the exact relationship between the optimal PSNE problem and the other problems.
For example the optimal PSNE problem is known to be tractable for singleton congestion games \cite{ieong2005fac} while we do not know how to solve the optimal CE problem. On the other hand for tree polymatrix games we showed the CE problem is in polynomial time, while the complexity of the PSNE problem is
unknown.
}

\fullver{
\begin{bf}Necessary condition for tractability.\end{bf}
Another open question is the following: is tractability of the deviation-adjusted social welfare problem a \emph{necessary} condition for tractability of the optimal CE problem?
We know (e.g., from \emcite{GLS1988}) that the separation oracle problem for the dual LP \eqref{dual} is equivalent to the problem of optimizing an arbitrary linear objective on the feasible set of \eqref{dual}.
However this in itself is not enough to prove equivalence of the deviation-adjusted social welfare problem and the optimal CE problem. First of all the
separation oracle problem is more general: it allows cutting planes other than constraints corresponding to pure strategy profiles.
Furthermore, \eqref{dual} has a particular objective, but optimizing an arbitrary linear objective means allowing
the objective to depend on $y$ as well as $t$. If we take the dual of such an LP with (e.g.) objective $r^Ty+t$ for some vector $r\in \Real^N$, we get a generalized version of
the optimal CE problem, with constraints $U x\geq r$ instead of $Ux\geq 0$.
}

\fullver{
\begin{bf}Relaxations and approximations.\end{bf}
Another interesting direction worth exploring is relaxations of the incentive constraints of these problems,
either as hard bounds or as soft constraints that add penalties to the objective, as well as the problem of approximating the optimal CE.
For these problems we can define corresponding variants of the deviation-adjusted social welfare problem as sufficient conditions, but it remains to be seen whether
one can prove concrete results, e.g., for approximating optimal CE for specific representations for which the exact optimal CE problem is hard.

\begin{bf}Communication complexity of uncoupled dynamics.\end{bf}
\emcite{hart2010com} considered a setting in which each player is informed only about her own utility function, and analyzed the communication complexity for so-called \emph{uncoupled} dynamics to reach various kinds of equilibrium. They used a straightforward adaptation of \emcite{PR08JACM}'s algorithm for a sample CE to show that a CE can be reached using polynomial amount of communication.
We can consider the question of reaching an optimal CE by uncoupled dynamics.
Our approach can be straightforwardly adapted to this setting, reducing the problem to finding a communication protocol
for the uncoupled version of the deviation-adjusted social welfare problem in which each player knows only her own utility function.
\begin{proposition}
If there is a polynomial communication protocol for the uncoupled
deviation-adjusted social welfare problem, then there is a polynomial communication protocol for the optimal CE problem.
\end{proposition}
At a high level, the protocol has a center running the ellipsoid method on \eqref{dual}, using  the communication protocol for the uncoupled
deviation-adjusted social welfare problem as a separation oracle.
An open problem is whether there exist more ``natural'' types of dynamics that converge to optimal CE.
For example, there is extensive literature on no-internal-regret learning dynamics that converges to the set of approximate CE in a polynomial number of steps.
Can such dynamics be modified to yield optimal CE?
} 

\hide{
\begin{bf}Representations of the set of CE.\end{bf}
\cite{PR08JACM}'s approach for reduced forms focused on an LP that expresses the incentive constraints and objectives
using marginal probabilities on the classes of the partitions in  the reduced form.
This LP is a relaxation of the optimal CE problem because the marginal probabilities do not always correspond to a distribution over strategy profiles.
However, if it is possible to add a polynomial number of constraints to the LP such that the
resulting set of constraints describes exactly the set of marginal probabilities that are extendable to distributions over strategy profiles, then
the optimum of the resulting LP corresponds to the optimal CE.
Such formulations are interesting because they provide a concise description of the set of CE, and thus may provide more information on the set of CE than
just the optimal CE.
\cite{PR08JACM} provided such a formulation for anonymous games and \cite{Kakade2003CEG}
for tree graphical games. For both cases there exist algorithms to  sample from the distributions described by such marginal probabilities.
Note that although such an approach is different from \cite{PR08JACM}'s main approach which reduced the problem to the separation problem,
anonymous games and tree graphical games have efficient algorithms for the separation problem as well as polynomial-sized descriptions of the set of CE.

Can such concise representations of the set of CE be constructed for game representations that are not captured by the reduced form framework?
In a recent paper \emcite{KXL11approxCE} gave such a construction for what they call graphical games with pair-wise utility functions,
which is equivalent to graphical polymatrix games. For tree graphs
the formulation is exact, while for general graphs it is a relaxation.
An open problem is whether this can be done for singleton congestion games.
It is also interesting to explore the relationship between the deviation-adjusted social welfare problem and the existence of concise representations of CE.
}

\nothesis{
\bibliographystyle{mlapa}
\shortver{\begin{footnotesize}}
\bibliography{Albert}
\shortver{\end{footnotesize}}
\fullver{
\appendix
\section{Proof of Theorem \ref{thm:optCE_scg}}\label{app:scg_proof}
\begin{proof}
As argued in Section \ref{sec:optCE_act_spec},
it is sufficient to construct a separation oracle for \eqref{cdual} that returns a player-symmetric cutting plane.
The cutting plane corresponding to a pure strategy profile solution $s$ of the coarse deviation-adjusted social welfare problem  is not player-symmetric in general; but we can symmetrize it by constructing a mixture of permutations of $s$. Since by symmetry each permuted version of $s$ correspond to a violated constraint, the resulting cutting plane is still correct and is symmetric. Enumerating all permutations over players would be exponential, but it turns out that for our purposes it is sufficient to use a small set of permutations.

Formally, let $\pi_i$ be the permutation over the set of players $\mathcal{N}$ that maps each $p$ to $p+ i \mod n$. Then the set  of permutations
$\{\pi_i\}_{0\leq i\leq n-1}$ corresponds to the cyclic group.

Suppose $s$ is a solution of the coarse deviation-adjusted social welfare problem with symmetric input $y$. The corresponding cut (violated constraint) is
$(C_s)^T y + w_s\leq t$. Recall that the $(p,j)$-th entry of $C_s$ is $C_s^{p,j}=(u^p_s-u^p_{js_{-p}})$.
For a permutation $\pi$ over $\mathcal{N}$, write $s^\pi$ the permuted profile induced by $\pi$, i.e. $s^\pi=(s_{\pi(1)},\ldots,s_{\pi(n)})$.
Then $s^\pi$ is also a solution of the coarse deviation-adjusted social welfare problem.
Form the following  convex combination of $n$ of the constraints of \eqref{cdual}:
\[
\frac{1}{n}\sum_{i=0}^{n-1} \left[(C_{s^{\pi_i}})^T y+ w_{s^{\pi_i}}\right] \leq t
\]
The left-hand side can be simplified to $w_s+(\overline{C}_s)^T y$ where $\overline{C}_s=\frac{1}{n}\sum_{i=0}^{n-1} C_{s^{\pi_i}}$.
We claim that this cutting plane is player-symmetric, meaning $\overline{C}_s^{p,j}=\overline{C}_s^{p',j}$ for all pairs of players $p,p'$ and all $j\in \mathcal{A}$.
This is because
\begin{align*}
\overline{C}_s^{p,j} & = \frac{1}{n}\sum_{i=0}^{n-1}C_{s^{\pi_i}}^{p,j}
= \frac{1}{n}\sum_{i=0}^{n-1} (u^p_{s^{\pi_i}}-u^p_{js^{\pi_i}_{-p}})\\
&= \frac{1}{n}\left[\sum_{\alpha\neq j} c(\alpha)f^\alpha(c(\alpha)) - (n-c(j))f^j(c(j)+1)\right]
=\overline{C}_s^{p',j}.
\end{align*}
This concludes the proof.
\nothesis{\qed}
\end{proof}
}
\end{document}